\documentclass [jap, aip, amsmath,amssymb,reprint]{revtex4-1}

\usepackage{color}
\usepackage{amssymb}
\usepackage{amsbsy}
\usepackage{amsmath}
\usepackage{amsfonts}
\usepackage{mathrsfs}
\usepackage{latexsym}
\usepackage[english]{babel}
\usepackage{url}
\usepackage[]{graphicx}
\usepackage{epstopdf}
\usepackage{subfigure}
\usepackage{hyperref}

\begin{document}
%
%
\title{Investigation of domain wall pinning by square anti-notches and its applications in three terminals MRAM}
%
\author{C. I. L. de Araujo}
\affiliation{Departamento de F\'{\i}sica, Laborat\'orio de Spintr\^onica e Nanomagnetismo, Universidade Federal de Vi\c{c}osa, Vi\c{c}osa, 36570-900, Minas Gerais, Brazil}
\author{J. C. S. Gomes}
\author{D. Toscano}  
\author{E. L. M. Paix\~ao}
\author{P. Z. Coura} 
\author{F. Sato}
\author{D. V. P. Massote}  
\author{S. A. Leonel} \email{sidiney@fisica.ufjf.br}
\affiliation{Departamento de F\'{\i}sica, Laborat\'orio de Simula\c{c}\~ao Computacional, Universidade Federal de Juiz de Fora, Juiz de Fora, Minas Gerais 36036-330, Brazil}
\date{\today}
\begin{abstract}
In this work we perform investigations of the competition between domain-wall pinning and attraction by anti-notches and finite device borders. The conditions for optimal geometries, which can attain a stable domain-wall pinning, are presented. This allow us the proposition of a three-terminals device based on domain-wall pinning. We obtain, with very small pulses of current applied parallel to the nanotrack, a fast motion of the domain-wall between anti-notches. In addition to this, a swift stabilization of the pinned domain-wall is observed with a high percentage of orthogonal magnetization, enabling high magnetoresistive signal measurement. Thus, our proposed device is a promising magnetoresistive random access memories with good scalability, duration, and high speed information storage.


\end{abstract}

\maketitle
\begin{figure}[htb!]
\centering
\includegraphics[width=6.2cm,clip]{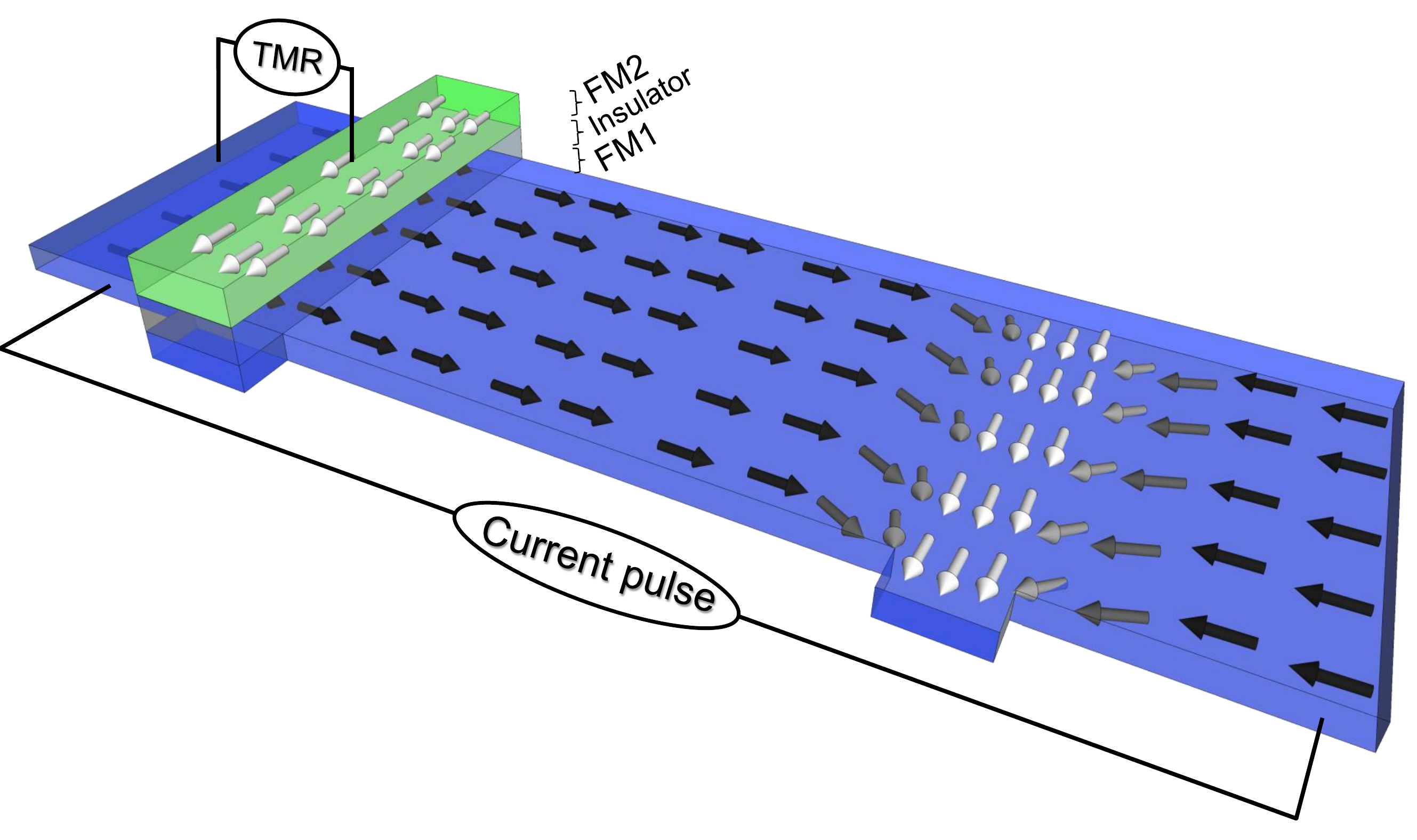}
\caption{Cartoon representing the proposed three terminal domain-wall based architecture. A current pulse in the short track moves the domain-wall between two anti-notches and the magnetoresistive signal is measured by the magnetic tunnel junction above one anti-notch.}
\label{figure1}
\end{figure}

The discovery of spin valve effects \cite{binasch1989enhanced,baibich1988giant,dieny1991giant} and magnetic tunnel junction measurements at room temperature \cite{moodera1995large,miyazaki1995giant}, allowed the development of several generations of magnetoresistive random access memories (MRAM) \cite{de2016multilevel}. A recent demonstration of MRAM integration among metallic contacts in silicon technology\cite{song2016highly} enables industrial large scale production and boosted further developments in scalability, consumption and speed. The MRAM generations can be divided according to the principle used for magnetization switching in the magnetic tunnel junction free layer. In the early generations, the magnetization switchings were made through Oersted fields generated by bit lines \cite{engel20054}, demanding large areas for the bit lines and high consumption due to the large currents needed. The next generation was developed with magnetization switching by spin transfer torque \cite{khvalkovskiy2013basic}. Such an approach represented a high gain in density, once there is no need of bit lines with switching performed by the current through the stack. However, the large current density needed can cause junction threshold, resulting in small durability. In order to protect the junction, the newest generations are based in three terminal devices with large currents passing by just the first ferromagnetic electrode and very small currents used to measure the tunnel magnetoresistance signal. Among such technology is the spin orbit torque MRAM \cite{cubukcu2014spin, garello2014ultrafast}, which uses heavy metals in the first layer to split the current into spin polarized channels, with high density enough to switch the first ferromagnetic layer by spin transfer torque. Another three terminal approach can be adapted from the original proposal of magnetic domain-wall based MRAM \cite{parkin2008magnetic}, which is based in domain-wall motion through a very long track and pinned by triangular notches, delimiting the bit length. Alternative geometries for bit length definition were also proposed \cite{al2016geometrically, goolaup2015transverse}. In this work, we investigate both, domain-wall attraction and pinning by square anti-notches, mapping best geometries for uniform pinned domain-wall, in order to measure stable, fast and highest values of tunnel magnetoresistive (TMR) signal by a magnetic tunnel junction (MTJ). The results, to be presented ahead, allowed the proposition of a three terminal domain-wall based MRAM, sketched in the cartoon presented in Figure \ref{figure1}. The working principle of such a device is based on a short current pulse applied in the device edges, in order to detach the domain-wall from the first anti-notch to be attracted by the second. Above the second anti-notch, a MTJ will act in the sensing, with the first ferromagnetic layer (FM1) where the domain-wall move, a thin insulator for the electronic tunneling and a second ferromagnetic layer (FM2), which is aligned orthogonal to the track magnetization by shape anisotropy. The tunnel magnetoresistive signal will vary from minimum to maximum, depending on the anti-notch where the domain-wall is pinned.      

In order to test the best geometry to achieve high performance in the proposed device, we have performed computational simulations. A Hamiltonian consisting of the isotropic Heisenberg model and the shape anisotropy can be used to describe a nanomagnet made of a soft ferromagnetic material:
%
\begin{multline}
H = J \Biggl \{ -  \sum_{<i,j>} \hat m_i \cdot \hat m_j \:+
\\
 + \frac{D}{J} \sum_{i,j} \biggl [\frac{\hat m_i \cdot \hat m_j - 3(\hat m_i  \cdot \hat{r}_{ij})(\hat m_j  \cdot \hat{r}_{ij})}{(r_{ij}/a)^3} \biggr ] \Biggr \}
\label{1}
\end{multline}
where $\hat m_i$ and $\hat m_j$ are unit vectors which represent the magnetic moments located at the $i$ and $j$ sites. The first term of equation (\ref{1}) describes the ferromagnetic coupling, whereas the second describes the the dipole-dipole interactions, which are responsible by the origin of the shape anisotropy.
In the micromagnetic approach, the renormalization of magnetic interaction constants depend not only on the parameters of the material, but also on the manner in which the system is partitioned into cells. According to the micromagnetic formulation, there is an upper limit for the work-cell size. Each micromagnetic cell hosts an effective magnetic moment $\vec m_i = (M_s V_{cel})\: \hat m_i$ aligned to the direction in which the atomic moments are saturated. From one cell to another, effective magnetic moments vary their directions gradually. These assumptions are only satisfied if we do not exceed the upper limit for the work-cell size. Therefore, the volume of the micromagnetic cell $V_{cel}$ has to be taken very carefully. In order to choose a suitable size for the work cell, we need to estimate characteristic lengths, which depend on the material parameters. For instance, the exchange length $\lambda = \sqrt{\frac{2A}{\mu_0 M_s^2}}$, provides an estimate of the exchange interaction range. In the simulations we have used typical parameters for Permalloy-79 (Ni$_{79}$Fe$_{21}$) with values as follow\cite{wysin_2010,Hertel_2007}: saturation magnetization $M_s = 8.6 \times 10^5$ A/m, exchange stiffness constant $A = 1.3 \times 10^{-11}$ J/m, and zero magnetocrystalline anisotropy. Thus, we have estimated $\lambda_{\mbox{\tiny{Py-79}}} \approx 5.3$ nm.
As in many micromagnetic simulation packages, we have used in our simulations the finite difference method, which subdivides the simulated geometry into cubic cells, that is, $V_{cel}=a^{3}$. In this context, the renormalization of the magnetic interaction constants are given by\cite{wysin_2010}: $J = 2\:a\:A$ and $\frac{D}{J} = \frac{1}{4 \pi} \left(\frac{a}{\lambda}\right)^2$. Based on the calculation of the exchange length for Permalloy-79, we have chosen the size of the micromagnetic cell as  $a=2\:\textrm{nm}< \lambda_{\mbox{\tiny{Py-79}}}$. Thus, planar nanowires have been spatially discretized into a cubic cell grid and the size of the work cell was chosen as $V_{cell} = 2 \times 2 \times 2$ nm$^3$, which is accurate enough for the current study.
%
%
%
%
The magnetization dynamics is governed by the Landau-Lifshitz-Gilbert (LLG) equation. In order to move the domain wall from one anti-notch to another, an electric current pulse is applied parallel to the nanotrack main axis. A generalized version of the LLG equation, which includes the spin torque effect has been proposed by Zhang and Li\cite{zhang2004s}. Thus, the domain wall dynamics driven by the spin-polarized current applied along the $x$-direction can be described by 
%
%
%
%
%
%
%
%
%
\begin{multline}
\frac{\partial \hat m_i}{\partial t'} = - \frac{1}{(1+\alpha^2)} \biggl \{ \hat m_i \times \vec b_i + \alpha \:\hat m_i \times \left( \hat m_i \times \vec b_i \right )+\\ + \frac{1}{(1+\beta^2)} \biggl ( \frac{u}{a\: \omega_0} \biggr ) \biggl [ \left (\beta -\alpha \right )\: \hat m_i \times \frac{\partial \hat m_i}{\partial x'} \:+\\ +  \left(1+\alpha \beta\right ) \: \hat m_i \times \left( \hat m_i \times \frac{\partial  \hat m_i}{\partial  x'}\right ) \biggl ] \biggr \}
\label{2}
\end{multline}
where the dimensionless effective field located at the micromagnetic cell $i$ is given by $\vec{b}_i = -J^{-1}\:\frac{\partial H}{\partial \hat{m}_i}$.
The first two terms take into account precession and damping torques, whereas the last two terms take into account the torque due to the injection of the spin-polarized electric current. The non-dimensional parameters, the Gilbert damping parameter $\alpha$ and the degree of non-adiabaticity $\beta$ are material parameters. Typical parameters for Permalloy-79 have been used in our simulations and the values are as follow~\cite{Hertel_2007,ratio_beta_alpha}: $\alpha = 0.01$ and $\beta = 0.015$. The influence of the ratio $\left(\beta / \alpha\right)$ on the dynamics of magnetic domain walls has already been investigated \cite{ratio_beta_alpha,beta_alpha,A_Thiaville_2005}.
%
%
The connection between the space-time coordinates and their dimensionless corresponding is given by: $\Delta x' = \Delta x / a$ and $\Delta t' = \omega_{0} \: \Delta t$, where $\omega_0 = \left(\frac{\lambda}{a}\right)^2 \gamma\: \mu_0  M_s$ is a scale factor with inverse time dimension, being $\gamma \approx 1.76\times 10^{11}\textrm{(T.s)}^{-1}$ the electron gyromagnetic ratio; for Permalloy $\mu_0  M_s \approx 1.0\:  \textrm{T}$. Thus, the product $(a\: \omega_0)$ has the dimension of distance divided by time (unit of velocity) as well as the term $u = j_e \left(\frac{g\:\mu_{\textrm{B}} }{2\:e M_s}\right)\:P $, where $j_e$ is the $x$-component of the electric current density vector (in our case, $\vec{j}_{e}=j_{e}\:\hat{x}$, so that $\vec{u}=u\:\hat{x}$ is a velocity vector directed along the direction of electron motion\cite{A_Thiaville_2005}). For Permalloy, the constant $\left(\frac{g\:\mu_{\textrm{B}} }{2\:e M_s}\right)\approx 6.7\times 10^{-11}\:\frac{\:\textrm{m}^3}{C}$, where $g$ is the Lande factor (for an electron $g\approx 2$), $\mu_{\textrm{B}}$ is the Bohr magneton and $e$ is the elementary positive charge. The non-dimensional parameter $P$ is the rate of the spin polarization. We used $P=0.5$, which amounts to those reported in Permalloy  nanowires of similar thicknesses\cite{P_Thickness}. 
%
%
%
%
%
%
We have implemented the fourth-order predictor-corrector method to solve numerically the equation (\ref{2}).
For Permalloy, the factor $\omega_0 \approx 1.33 \times 10^{12}\: \textrm{s}^{-1}$. Thus, the time step $\Delta t' = 0.01$ used in the numerical simulations corresponds to $\Delta t \approx 7.5 \times 10^{-15} \:\textrm{s}$. In micromagnetic simulations, we have used our own computational code, which has been used in several works of our group\cite{paixao2018depinning,nosso3}.

%
%
\begin{figure*}[htb!]
\centering
\includegraphics[width=13cm,clip]{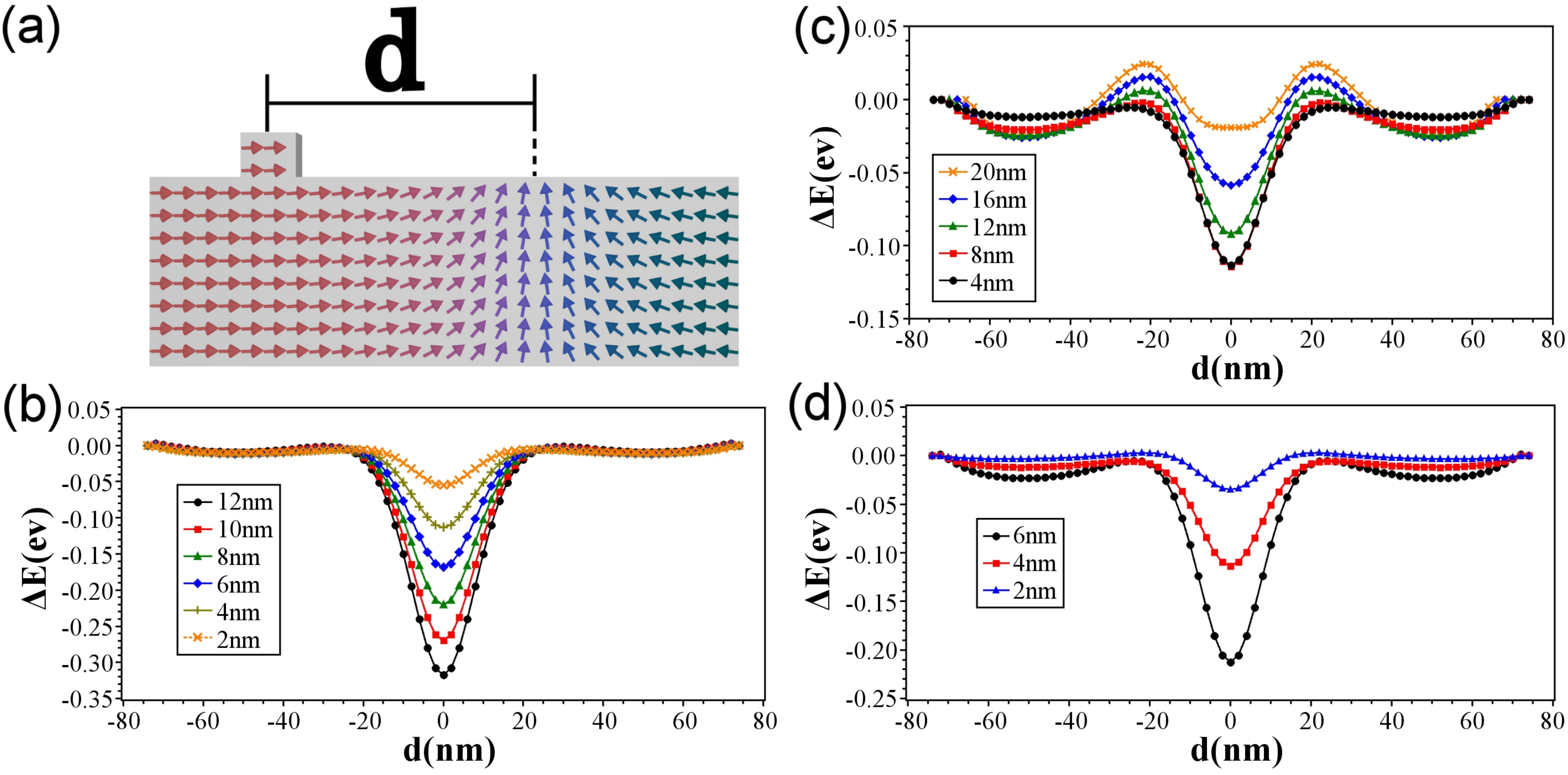}
\caption{(a) Schematic view of how the distance $d$ between the center of the anti-notch and the center of the TDW in the nanotrack was considered. The color gradient in the arrows represents the magnetic moment's directions. We have analyzed the interaction energy as a function of the distance between the center of the anti-notch and the center of the TDW by varying (b) anti-notch width ($W_{not}$), considering constant anti-notch length $L_{not} = 4$ nm and thickness $T = 4$ nm, (c) anti-notch length ($L_{not}$), considering constant anti-notch width $W_{not} = 4$ nm and thickness $T = 4$ nm and (d) anti-notch thickness $T$, considering constant anti-notch length $L_{not} = 4$ nm and  width $W_{not} = 4$ nm.}
\label{d_notch}
\end{figure*}

In the simulations, we have considered the Permalloy planar nanowires with length $L = 152$ nm and width $W = 16$ nm. The anti-notch thickness is the same as that of the nanotrack thickness $T$. The anti-notch parameters, that is, the anti-notch length $L_{not}$, as well as the anti-notch width $W_{not}$ were varied throughout the study.
 The stopping criterion for the relaxation consists of integrating the LLG equation without an external agent (magnetic field or spin-polarized current) until both the energy of the system and its magnetization vector stop oscillating. Thus, the system reaches the equilibrium magnetic state, which provides the possibility of the adjustment of the TDW width\cite{toscano2014position}. 
See \textcolor{blue}{supplementary material} for details of the relaxation simulations. The equilibrium configuration obtained in this way has been used as the initial configuration in other simulations where a single anti-notch was inserted into the nanowire. To calculate the interaction energy $\Delta$E between the TDW and the anti-notch as a function of the center-to-center separation $d$, we fix the TDW at the center of the nanowire and varied only the anti-notch position along the nanowire edge, see Fig. \ref{d_notch}(a). For each separation $\textrm{d}$, the total energy of the system is calculated using Eq.(\ref{1}), and the interaction energy has been estimated using the following expression: $\Delta$E$_i$ = E$_i$ - E$_0$, where E$_i$ represents the total energy of the nanowire that hosts the anti-notch at any position $x$, whereas E$_0$ is the reference energy, in which the anti-notch is located at the maximum possible distance from the wall, that is, at the corner of the nanowire. Figures \ref{d_notch}(b), \ref{d_notch}(c) and \ref{d_notch}(d), show the behavior of the interaction energy as a function of the distance $d$ between the center of the anti-notch and the center of the TDW, as we vary the anti-notch parameters. It can be observed from Figure \ref{d_notch} that the anti-notches work as pinning traps for the TDW and the interaction strength increases as we increase the anti-notch width ($W_{not}$) and thickness ($T$), but decreases as we increase the anti-notch length ($L_{not}$).

\begin{figure}[htb!]
\centering
\includegraphics[width=7cm,clip]{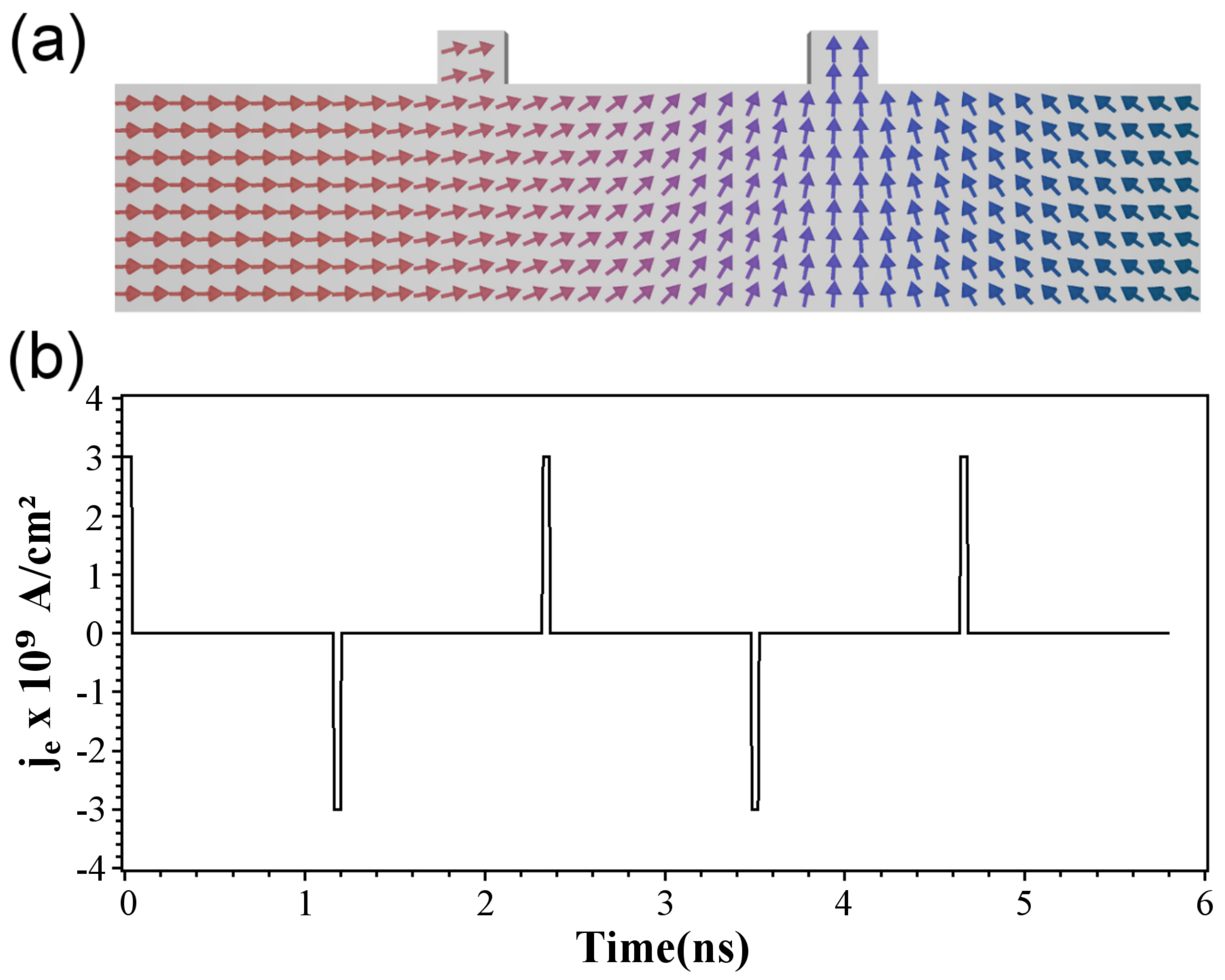}
\caption{(a) Schematic view of a nanotrack containing two anti-notches. The anti-notch on the right side has magnetization perpendicularly aligned to the easy axis. The anti-notch on the left side should exhibit magnetization parallel to the easy axis, but due to its proximity to the other anti-notch, its magnetization is aligned in an intermediate direction. (b) Sequence of spin-polarized current pulses applied along the $x$-axis to move the TDW from one anti-notch to another (Multimedia view).}
\label{xnotch}
\end{figure}
From now on, we consider two identical anti-notches equidistant from the nanowire width axis.
%
Based on our observations, we choose a nanotrack with thickness $T = 4$ nm, containing a pair of square anti-notches  $L_{not} = W_{not} = 4$ nm in order to investigate the TDW magnetization as a function of the relative distance between the anti-notches $x_{not}$. The logic states (“0” and “1”) are defined according to the anti-notch magnetization, if it is aligned parallel or perpendicular to the nanotrack easy axis. Therefore, any intermediate direction would hinder the information reading in the device, decreasing the TMR signal. In some of the tested configurations, after the system reaches the relaxed magnetic state in which the TDW was located near the anti-notch on the right, we observed that due to the proximity between anti-notches, the  magnetization of the anti-notch on the left  was aligned with an intermediate direction between parallel and perpendicular directions, as can be seen in the Figure \ref{xnotch}(a) (Multimedia view). However, when increasing the separation between the anti-notches at $x_{not} = 18$ nm, we were able to achieve at least $99.5\:\%$ of magnetization aligned parallel to the easy axis for the anti-notch on the left.
Additionally, we consider a variety of possible candidates for the storage cells of the random access memory proposed in this paper. Using the initial condition of the wall close to the anti-notch on the right, we have numerically calculated the relaxed micromagnetic state of several nanowires with different parameters of the anti-notch arrangement. From these equilibrium magnetic configurations, we applied a sequence of current pulses ($j_e = \pm \:3 \times 10^9$ A/cm$^2$ with duration of $\Delta t \approx 0.04$ ns) separated by a time interval of relaxation ($j_e = 0$ with duration of $\Delta t \approx 1.12$ ns) in order to move the wall from one anti-notch to another as shown in 	Figure \ref{xnotch}(b).
\begin{figure}[htb!]
\centering
\includegraphics[width=7.0cm,clip]{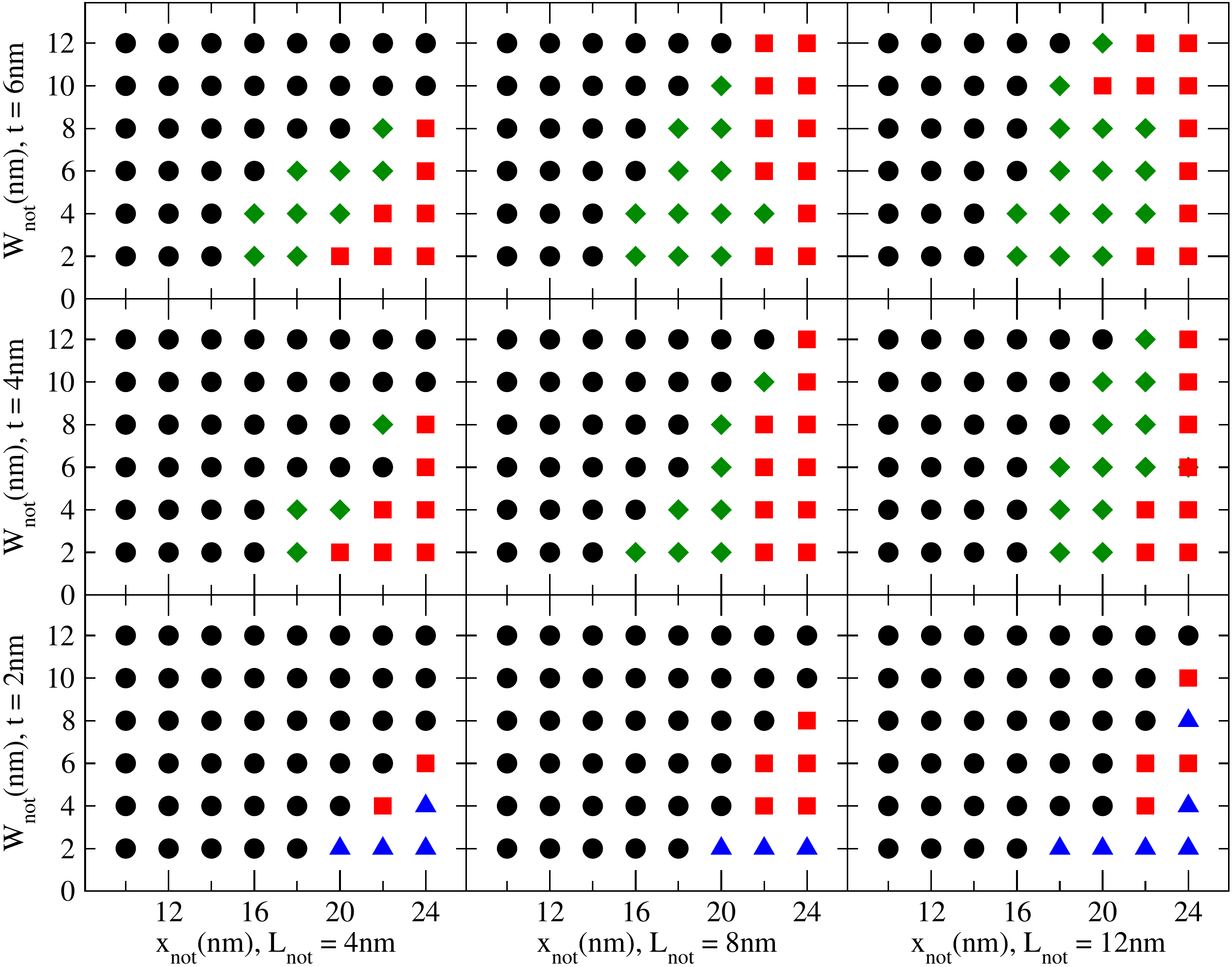}
\caption{TDW position and  anti-notch magnetization controllability diagrams, which summarize micromagnetic simulation results of a single TDW in a Permalloy planar nanowire with two identical anti-notches. Before applying the sequence of current pulses, we checked if the TDW was really pinned at the anti-notch on the right. Although the anti-notches work as pinning traps, their pinning potential strength cannot be strong enough to pin the wall, so that the TDW is expelled through one of the nanowire ends. Relaxation results in which the TDW was expelled through the right side of the nanowire are represented by blue triangles. Relaxation results in which the TDW was pinned at the anti-notch on the right are represented by black circles, however, the magnetization of the left side anti-notch was not aligned with the magnetization easy axis of the nanowire, such as is shown in Fig. \ref{xnotch}(a). Red squares correspond to the simulation results in which the TDW was expelled from the nanowire, after the application of the current pulse sequence. Green diamonds correspond to the simulation results in which we observed the TDW position accurate control, that is, not only the TDW position could be controlled from one anti-notch to another, but also the magnetization vectors of anti-notches did present parallel (TDW absence) and perpendicular (TDW presence) alignments with the magnetization easy axis of the nanowire, before the next current pulse is applied.}
%
%
\label{diagram2}
\end{figure}

Over a wide range of the  anti-notch parameters and the spacing between anti-notches, we numerically calculated the dynamic response of the wall under the influence of the above-mentioned current pulse sequence. The simulation results have been organized into event diagrams (see Fig. \ref{diagram2}), which show magnetization configuration of the nanowire before and after the application of the current pulse sequence.
Analyzing Fig. \ref{diagram2}, it can be noted that the precise control of the TDW position is only possible when the geometric factors of the anti-notches are adjusted properly, such as the spacing between them and the parameters of the spin-polarized current pulse simultaneously.  
%
%
%
%
%
%
%
%
%
\begin{figure}[htb!]
\centering
\includegraphics[width=8.6cm,clip]{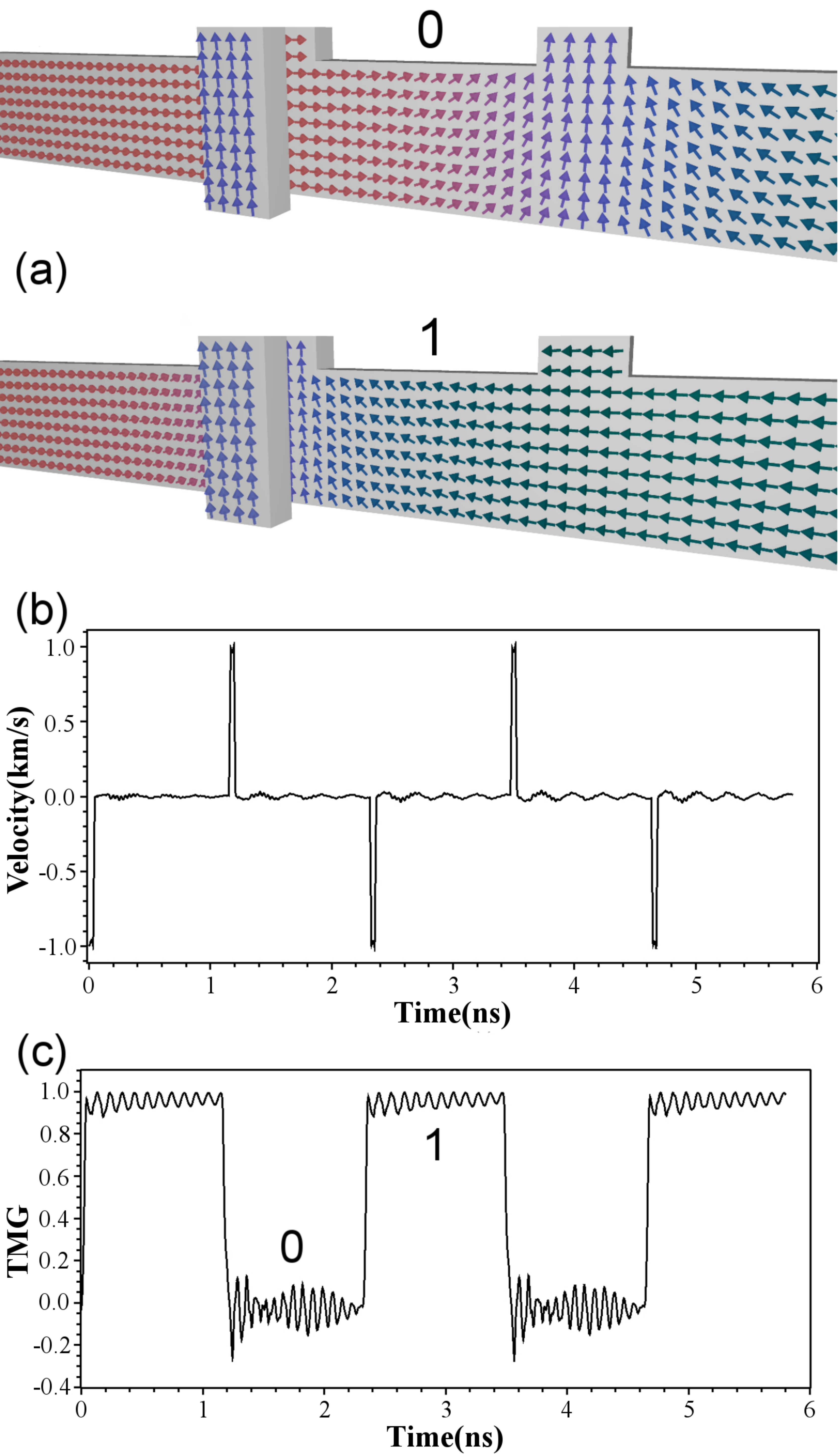}
\caption{(a) Schematic view in perspective of the possible states of layers magnetization with state in which the TDW is at the notch far from the reference layer ``0'' (before the application of the first current pulse) and in the state in which the TDW is at the notch near the reference layer ``1'' (shortly after the application of the first current pulse). (b) and (c) Time evolution of the TDW velocity and $TMG$, respectively (Multimedia view).}
\label{layer}
\end{figure}

%
We fixed the distance between layers $d_{lay} = 2$ nm, which is a good approximation for average thickness used in general for MTJ. Due to shape anisotropy, the reference layer magnetization always remains aligned parallel to the easy axis of this layer. The anti-notches are inserted into the storage-layer and their two logic states (``0'' and ``1'') corresponding to two possible magnetization orientations. The MTJ is connected to a selection transistor and, upon reading, a small electric current flows through the MTJ. The information bit would correspond to the MTJ resistance. 
Figure \ref{layer}(a) (\textcolor{blue}{Multimedia view})  shows the TDW at the notch far from the reference layer (before the application of the first current pulse) and at the notch near the reference layer (shortly after the application of the first current pulse). During the current pulse, the TDW reaches a velocity of approximately $1\:\textrm{km/s}$, as shown in Figure \ref{layer}(b), with velocity $v(t) = \frac{L}{2} \frac{d<M_x (t)>}{dt}$, where $M_x$ is the $x$-component of the system magnetization vector. The calculation of the domain wall velocity has been previously proposed \cite{DW_velocity}. The local conductance, that is just the inverse of the resistance, is given by the scalar product of the facing magnetic moments on both sides of the tunnel barrier $TMG= \frac{\sum \hat m_i \cdot \hat m_j}{n}$, where $\hat m_i$ and $\hat m_j$ are the facing magnetic moments on the storage and the reference layers, respectively, and $n$ is the total number of magnetic moments in the layer \cite{de2016multilevel}. 
The calculated $TMG$ evolution to the considered configuration, presented in the Figure \ref{layer} (c), shows variation between 0 and 1 presenting negligible signal fluctuations due to a small magnetization oscillation during the change of states. Due to shape anisotropy, the magnetization in the reference layer remains aligned to the major axis. The reference layer major axis direction is perpendicular to the recording layer easy axis, but it is parallel to the domain wall magnetization direction. Thus, when the wall reaches the anti-notch having the reference layer (state ``1''), the interaction between the reference layer and the domain wall favors the parallel alignment of its magnetizations, decreasing the fluctuations.

In conclusion, we have mapped the conditions for domain-wall pinning with or without current pulse applied in function of a set of anti-notch parameters. In addition, we found an optimal geometry, as small as the dimensions used in several MRAM investigated in literature. In the investigated geometry, we observed a swift domain wall motion between anti-notches with short current pulse with a duration of $\Delta t\approx 0.04$ ns. The current used is similar to the ones already used in another investigated devices \cite{yang2015domain, loreto2018creation} which demonstrates that it would not characterize any damage to a device in such a short operational time. The observed stable pinning and magnetization stabilization in $\Delta t\approx 1.12$ ns allow quite fast information storage, compared to a fast MRAM described in literature \cite{cubukcu2018ultra} and the high percentage of uniformity in the orthogonal magnetization of domain-wall pinned in the anti-notch enable maximum TMR to be measured by the MTJ.

\vspace*{0.3cm}

See \textcolor{blue}{supplementary material}, in which we describe how to obtain equilibrium magnetic states and provide our stopping criterion for the relaxation micromagnetic simulations. In particular, we give an example to obtain the relaxed micromagnetic state of a single transverse domain wall in a Permalloy planar nanowire.

\vspace*{0.3cm}

This study was financially supported in part by the Coordena\c{c}\~ao de Aperfei\c{c}oamento de Pessoal de N\'{\i}vel Superior - Brasil (CAPES) - Finance Code 001 and also supported by CNPq and FAPEMIG (Brazilian agencies). We greatfully thank to our friend Saif Ullah for making the English revision of this paper.


\begin{thebibliography}{100}
\bibitem{binasch1989enhanced}{G. Binasch, P. Gr{\"u}nberg, F. Saurenbach, and W. Zinn, Phys. Rev. B {\bf 39}, 4828-4830 (1989).}
\bibitem{baibich1988giant}{M. N. Baibich, J. M. Broto, A. Fert, F. Nguyen Van Dau, F. Petroff, P. Etienne, G. Creuzet, A. Friederich, and J. Chazelas, Phys. Rev. Lett. {\bf 61}, 2472-2475 (1988).}
\bibitem{dieny1991giant}{B. Dieny, V. S. Speriosu, S. S. P. Parkin, B. A. Gurney, D. R. Wilhoit, and D. Mauri, Phys. Rev. B {\bf 43}, 1297-1300 (1991).}
\bibitem{moodera1995large}{J. S. Moodera, Lisa R. Kinder, Terrilyn M. Wong, and R. Meservey, Phys. Rev. Lett. {\bf 74}, 3273-3276 (1995).}
\bibitem{miyazaki1995giant}{T. Miyazaki and N. Tezuka, J. Magn. Magn. Mater. {\bf 139}, 231-234 (1995).}
\bibitem{de2016multilevel}{C. I. L. de Araujo, S. G. Alves, L. D. Buda-Prejbeanu, and B. Dieny, Phys. Rev. Applied {\bf 6}, 024015 (2016).}
\bibitem{song2016highly}{Y. J. Song, J. H. Lee, H. C. Shin, K. H. Lee, K. Suh, J. R. Kang, S. S. Pyo, H. T. Jung, S. H. Hwang, G. H. Koh et al., IEEE International Electron Devices Meeting (IEDM) {\bf 27.2.1-27.2.4}, 663-666 (2016).}
%
\bibitem{engel20054}{B. N. Engel, J. \AA kerman, B. Butcher, R. W. Dave, M. DeHerrera, M. Durlam, G. Grynkewich, J. Janesky, S. V. Pietambaram, N. D. Rizzo et al., IEEE Trans. Magn. {\bf 41}, 132-136 (2005).}
%
\bibitem{khvalkovskiy2013basic}{A. V. Khvalkovskiy, D. Apalkov, S. Watts, R. Chepulskii, R. S. Beach, A. Ong, X. Tang, A. Driskill-Smith, W. H. Butler, P. B. Visscher et al., J. Phys. D: Appl. Phys. {\bf 46}, 074001 (2013).}
\bibitem{cubukcu2014spin}{M. Cubukcu, O. Boulle, M. Drouard, K. Garello, C. O. Avci, I. M. Miron, J. Langer, B. Ocker, P. Gambardella, and G. Gaudin, Appl. Phys. Lett. {\bf 104}, 042406 (2014).}
\bibitem{garello2014ultrafast}{K. Garello, C. O. Avci, I. M. Miron, M. Baumgartner, A. Ghosh, S. Auffret, O. Boulle, G. Gaudin, and P. Gambardella, Appl. Phys. Lett. {\bf 105}, 212402 (2014).}
\bibitem{parkin2008magnetic}{S. S. P. Parkin, M. Hayashi, and L. Thomas, Science {\bf 320}, 190-194 (2008).}
%
\bibitem{al2016geometrically}{M. Al Bahri and R. Sbiaa, Sci. Rep. {\bf 6}, 28590 (2016).}
\bibitem{goolaup2015transverse}{S. Goolaup, M. Ramu, C. Murapaka, and W. S. Lew, Sci. Rep. {\bf 5}, 9603 (2015).}



\bibitem{wysin_2010} {G. M. Wysin, J. Phys.: Condens. Matter {\bf 22}, 376002 (2010).}


\bibitem{Hertel_2007}{R. Hertel, S. Gliga, M. F\"{a}hnle, and C. M. Schneider, Phys. Rev. Lett. {\bf 98}, 117201 (2007).}


%







\bibitem{zhang2004s}{S. Zhang and Z. Li, Phys. Rev. Lett. {\bf 93}, 127204 (2004).}








\bibitem{ratio_beta_alpha}{H. Y. Yuan and X. R. Wang, Phys. Rev. B {\bf 92}, 054419 (2015).}




\bibitem{beta_alpha}{Y. Tserkovnyak, A. Brataas, and G. E. W. Bauer, J. Magn. Magn. Mater. {\bf 320}, 1282-1292 (2008).}







\bibitem{A_Thiaville_2005} {A. Thiaville, Y. Nakatani, J. Miltat, and Y. Suzuki, Europhys. Lett. {\bf 69}, 990-996 (2005).}











\bibitem{P_Thickness}{M. Haidar and M. Bailleul, Phys. Rev. B {\bf 88}, 054417 (2013).}








%



%
%

\bibitem{paixao2018depinning}{E. L. M. Paix\~{a}o, D. Toscano, J. C. S. Gomes, M. G. Monteiro Jr., F. Sato, S. A. Leonel, and P. Z. Coura, J. Magn. Magn. Mater. {\bf 451}, 639-646 (2018).}

\bibitem{nosso3} {D. Toscano, S. A. Leonel, P. Z. Coura, F. Sato, B. V. Costa, M. V\'{a}zquez, J. Magn. Magn. Mater. {\bf 419} (2016) 37-42.}


\bibitem{toscano2014position}{D. Toscano, V. A. Ferreira, S. A. Leonel, P. Z. Coura, F. Sato, R. A. Dias, and B. V. Costa, J. Appl. Phys. {\bf 115}, 163906 (2014).}
%


\bibitem{DW_velocity} {D. G. Porter and M. J. Donahue, J. Appl. Phys. {\bf 95}, 6729 (2004)}.








\bibitem{yang2015domain}{See-Hun Yang, Kwang-Su Ryu, and Stuart Parkin, Nat. Nanotechnol. {\bf 10}, 221-226 (2015).}
\bibitem{loreto2018creation}{R. P. Loreto, W. A. Moura-Melo, A. R. Pereira, X. Zhang, Y. Zhou, M. Ezawa, and C. I. L. de Araujo, J. Magn. Magn. Mater. {\bf 455}, 25-31 (2018).}
\bibitem{cubukcu2018ultra}{M. Cubukcu, O. Boulle, N. Mikuszeit, C. Hamelin, T. Br\"{a}cher, N. Lamard, M.-C. Cyrille, L. Buda-Prejbeanu, K. Garello, I. M. Miron et al., IEEE Trans. Magn. {\bf 54}, 4 (2018).}





%
%
%
%
%
%






 
\end{thebibliography}


%
\end{document}